\documentstyle [pre,aps,preprint]{revtex}
\newcommand{\ro}{\mbox{\boldmath $\rho$}}
\newcommand{\rv}{\vec{r}}
\newcommand{\vv}{\vec{v}}
\newcommand{\tpart}{\tilde{\partial}}

\begin{document}
\draft
\title{Chaotic Properties of Dilute Two and Three Dimensional Random Lorentz
Gases II: Open Systems}
\author{H. van Beijeren} 
\address{Institute for Theoretical Physics, University of Utrecht, \\
Postbus 80006, Utrecht 3508 TA, The Netherlands} 
\author{Arnulf Latz} 
\address{Johannes Gutenberg-Universit\"at, 
Institut f\"ur  Physik, \\  
55099 Mainz, Germany}
\author{J. R. Dorfman} 
\address{Institute for Physical Science and Technology, and
Department of Physics, \\  
University of Maryland, College Park, Maryland,
20742, USA}
\date{\today}
\maketitle

\begin{abstract}
We calculate the
spectrum of Lyapunov exponents for a point particle moving in a random
array of fixed hard disk or hard sphere scatterers, i.e.
the
disordered 
Lorentz gas, in a generic nonequilibrium situation. In a
large system which is finite in at least some directions, and with
absorbing boundary conditions, the  
moving particle escapes the system with probability one.  However,
there is a set of zero Lebesgue measure of
initial phase points for
 the moving particle, such that escape never occurs. Typically, this
set of points forms a   
fractal repeller, and the Lyapunov spectrum is calculated here for
trajectories on this repeller.  
For this calculation, we need the solution of the
recently introduced extended Boltzmann equation for the
nonequilibrium distribution of the radius of curvature matrix and the
solution of the standard Boltzmann equation. 
The escape-rate formalism then gives an explicit result for the 
Kolmogorov Sinai entropy on the repeller.

\end{abstract}

\section{Introduction}
In this paper we extend the analysis of the chaotic properties of
dilute, random Lorentz gases given in \cite{vbld} (denoted by (I)) to
include open systems with absorbing boundaries. The Lorentz gas
consists of a point particle moving in a system of identical hard disk
(d=2) or hard sphere (d=3) scatterers of radius $a$. In a dilute,
random Lorentz gas, the average distance between the
scatterers is large compared to their radius, $a$, and the scatterers
are placed at random on the plane or in space without overlapping each
other.
The interest in open systems, with absorbing boundaries, is occasioned
by the escape-rate method of Gaspard and Nicolis \cite{gn90} which
relates the coefficient of diffusion for the moving particle in the
Lorentz gas to the dynamical properties of particles on the set of
trajectories that never escape from the system.

 The method of Gaspard and Nicolis is based on an identity in the
theory of open, hyperbolic dynamical systems, called the escape-rate
formula. This formula is an expression for the rate of decay of the
probability, $P(t)$, of finding a moving particle in an open region,
$V$, surrounded by absorbing boundaries, at time $t$. If the motion of
the moving particle is  hyperbolic, the time dependence of $P(t)$ is
exponential, decaying as $\exp(-\gamma t)$,  where the escape-rate,
$\gamma$, is given by 
\begin{equation}
\gamma = \sum_{\lambda_{i}>0}\lambda_{i}({\cal R}) - h_{KS}({\cal{R}}).
\label{er1}
\end{equation}
Here ${\cal{R}}$ denotes the set of initial phase points for the
moving particle which are on  trajectories that never escape from the
system. This set of points is called a ``repeller'' in
the phase space, typically of measure zero with respect to the usual
Lebesgue measure, with unstable and  stable manifolds characterized by
Lyapunov exponents $\lambda_{i}({\cal{R}})$,  and Kolmogorov-Sinai
(KS) entropy $h_{KS}({\cal{R}})$. The sum in Eq. (\ref{er1})  is only
over the positive Lyapunov exponents,  and the Lyapunov exponents and
KS entropy are to be calculated with respect  to an appropriate measure on the repeller \cite{bohrrand,gasdo}.  This equation may be
considered as the microscopic expression for the escape-rate of the
particle from the open region, $V$. A macroscopic expression  for the
escape-rate is provided by the diffusion equation satisfied by $P(t)$
on large time and large space scales,
\begin{equation}
\frac{\partial P(t)}{\partial t} = D \nabla^{2}P(t),
\label{er2}
\end{equation}
where D is the macroscopic diffusion coefficient for the moving particle in
$V$. The solution  of this equation, for long times, and for absorbing
boundary conditions is of the form 
\begin{equation}
P(t)= \exp[-(\frac{
c}{L^{2}})D t],
\label{er3}
\end{equation}
where $L$ is a length characterizing the distance to the absorbing
boundary of interior points in $V$, and 
c is a numerical factor
determined by the
shape of $V$ and the absorbing boundary conditions. Since the
microscopic and the macroscopic expressions for the escape-rate describe
the same escape process, they have to be identical, which leads to the Gaspard-Nicolis formula 
\begin{equation}
D = \lim_{L\rightarrow\infty}\frac{L^{2}}{
c}\left[
\sum_{\lambda_{i}>0}\lambda_{i}({\cal R}) - h_{KS}({\cal{R}})\right].
\label{er4}
\end{equation}
Here the limit $L\rightarrow\infty$ is taken to remove finite size
corrections, including the effects of microscopic boundary layers. 
It is worth mentioning that the Green-Kubo microscopic expressions for the diffusion
coefficient, $D$, and Sinai's expression for the sum of positive
Lyapunov exponents represent these quantities as infinite time
integrals over appropriate dynamical quantities. In order to apply
Eqs. (\ref{er2})-(\ref{er4}), without having to worry about subtleties
due to the possible slow convergence of these integrals (due to long-time-tail effects)
we assume that all of the dynamical
quantities, $\lambda_{i}, h_{KS}$ reach their long-time asymptotic
values on a time scale which is short compared to  the hydrodynamic
time scale on which the diffusion equation applies\footnote{This
assumption is certainly a reasonable one for the low density cases
we consider here.}. The 
escape-rate formula for diffusion, Eq. (\ref{er4}), and the
generalizations to other  transport coefficients \cite{dorgas}, show a
striking connection between the macroscopic  quantities that control
hydrodynamic processes, the transport coefficients, and  
the microscopic quantities that describe the chaotic dynamics taking
place on the repeller. A detailed discussion of this can be
found elsewhere \cite{gasbook,jrdbook}. 

The purpose of this paper is to provide an analytical calculation of
the positive Lyapunov 
exponents on the repeller, $\lambda_{i}({\cal{R}})$, for the random,
dilute Lorentz gas, and to use these, together with known values for
the diffusion coefficient, $D$, to determine $h_{KS}({\cal{R}})$, the
KS entropy of the trajectories on the repeller. We take the same
approach as in I where we used kinetic theory arguments to calculate
the Lyapunov spectra of two and three dimensional random, dilute
Lorentz gases in equilibrium. Here we are concerned with a
nonequilibrium situation where particles escape from the
system. 
We
will see that spatial inhomogeneities introduced by the absorbing
boundaries will require some significant modifications to our previous
calculations. There we used mean-free-path arguments and some results
from the theory of products of random matrices to determine low
density values for the individual Lyapunov exponents and we used an
extended Lorentz-Boltzmann equation as an efficient method for
determining the sum of the positive Lyapunov exponents. Here we will
do the same for the Lyapunov exponents on the repeller. We mention
that before we developed this method \cite{vbd} analytical results for Lyapunov
exponents on repellers had only been obtained for simple one
dimensional models \cite{gaspholian,ott}. Otherwise one had to use
numerical methods \cite{gaspbaras}.

It will be helpful to recall some ideas from I. There we obtained the
individual Lyapunov  exponents as well as their sums in terms of
various averages over functions  of a radius of curvature
(ROC) matrix
$\ro$, using an appropriate distribution function.   A central
notion introduced in I is the use of an extended Lorentz-Boltzmann
equation (ELBE) to determine the distribution of the elements of the
radius of curvature matrices needed for our calculations.
The ELBE was derived heuristically in I. In the
absence of external fields  acting on the moving particle it is given by
\begin{eqnarray}
&& \frac{\partial F}{\partial t}+\vv\cdot\nabla F +\dot{\ro}:
\frac{\partial F}{\partial \ro} \nonumber \\ & = & na^{d-1}\int
d\hat{n}|\vv\cdot\hat{n}|\left[\Theta(\vv\cdot\hat{n})\int d \ro'\delta(\ro
-\ro'(\ro))F(\rv,\vv\prime,\ro',t)- 
\Theta(-\vv\cdot\hat{n})F(\rv,\vv,\ro,t)\right].
\label{er5}
\end{eqnarray}
The notation is exactly the same as in I. The primed
variables denote the restituting values, which lead to the unprimed
values after a collision, and $\hat{n}$ denotes a unit vector from the
center of a scatterer to the point of impact of the moving particle at
a collision. The solution of Eq. (\ref{er5}) is
normalized according to
\begin{equation}\label{3.2}
\int d \ro' F(\rv,\vv,\ro',t) =  f_B(\rv, \vv,t), 
\end{equation} 
where $f_B$ is the solution of the standard Lorentz--Boltzmann equation\cite{chapcow}.
In this paper we
will assume, as in I, that the
elements of the inverse of the restituting matrix, $[\ro']^{-1}$, are
typically small compared to the inverse of the
scatterer radius $a^{-1}$. 
As a result, we may simplify the delta function appearing in the collision
integral on the right hand side of Eq. (\ref{er5}).

The plan of the paper is as follows: In Section II we discuss the
problem of calculating averages on the repeller in phase space. There
we argue that it is necessary to introduce a survival probability
for a particle that is inside the system at time $t$, to be still 
inside the system at time $t+T$. We argue that this probability is
needed to guarantee that the properties we calculate are actually
those of the repeller and not merely some asymptotic properties of a
set of particles that eventually escape from the region $A$. (Not taking this survival probability into account led to an
erroneous result in a previous publication on this subject,
subsequently corrected in an erratum \cite{vbd}.)  We will set up the
formalism that will allow us to calculate the Lyapunov spectrum on the
repeller in Section III. In Section IV, we will treat the two dimensional
case, and in Section V, the three dimensional case.  We conclude in
Section VI with a  summary of our results, a number of remarks,  and a
discussion of some interesting open questions.

\section{Averaging on the repeller}

To treat the open system correctly it is necessary to develop a tool
which guarantees that the quantities we calculate are the actual
Lyapunov exponents for trajectories on the repeller. 
In general, the measure on a repeller is a very singular object. For large 
systems
we expect that this measure will be very similar for all typical configurations of scatterers,
when observed on length scales small compared to typical macroscopic length scales (system size)
but large compared to the mean free path between collisions. Therefore averages on the repeller may be replaced by averages over smooth reduced distribution functions for the light particle alone obtained by 
averaging over all configurations of scatterers.
Further support of this picture is provided by the observation that for
large systems the fractal dimension is close
to the embedding dimension, as follows from  
Young's formula \cite{ott} relating the
Lyapunov exponents to the information dimension of the repeller.

For dilute Lorentz gases without escape, the probability density of the
ROC matrix elements is obtained as the time independent solution of the generalized
Lorentz-Boltzmann equation for $F(\ro)$ given by Eq. (\ref{er4}). For a system
with escape, the solution of the time dependent
Lorentz-Boltzmann equation only determines the probability of finding a
particle at point $\vec{r}$ with velocity $\vec{v}$ with a ROC matrix
$\ro$ at time t. It does not contain any information about the future
behavior of this particle. More specifically it does not exclude the
possibility of this particle leaving the system at a later time $t+\tau$. To
obtain the smoothed density {\em on the repeller} we therefore have to
weight this Lorentz-Boltzmann density with the survival probability. This is
the conditional probability $S(\rv,\vv,t|t+T)$, that a particle at
point $\rv$ with velocity $\vv$  at time t will still be in the
system at the time $t+T$. It does not depend on $\ro$ because all
trajectories in a bundle are infinitesimally close to each other. The introduction of the survival probability
in order to obtain a proper description of the repeller is very
reminiscent of, and essentially identical to, methods used to compute
fractal dimensions and other properties of attractors and repellers in
more traditional dynamical systems calculations \cite{ott,huntpre}.

The survival probability that we need, $S(\rv,\vv,t|t+T)$, can, in d
dimensions, be written as an 
integral over
the conditional probability $S(\rv,\vv,t|\rv\prime,\vv\prime,t+T)$ as

\begin{equation} 
S(\rv,\vv,t|t+T)  =  \int d^d \rv\prime d^d \vv\prime S(\rv,\vv,t|\rv\prime,\vv\prime,t+T),
\label{survival}
\end{equation}
with initial condition
\begin{equation}
S(\rv,\vv,t|\rv',\vv',t)  =  \delta(\rv -\rv') \delta(\vv-\vv')
\label{initial}
\end{equation}

The average of a function $g(\ro)$ on the repeller is then given by 

\begin{equation} \label{repelav}
\langle g(\ro) \rangle_{Rep} = \lim_{t \to \infty} \lim_{T \to \infty}
\frac{\displaystyle \int d \ro 
d \rv d \vv F(\rv, \vv, \ro,t) S(\rv,\vv,t|t+T) g(\ro)}{\displaystyle
\int d \ro d \rv d \vv F(\rv, \vv, \ro,t) S(\rv,\vv,t|t+T)}
\end{equation} 

The limit $T \to \infty$ has to be taken first, to guarantee that only
the trajectories which never leave the system are counted. The limits
of numerator and denominator vanish separately, since the repeller is
a set of Lebesgue measure zero. The conditional probability
$S(\rv,\vv,t|\rv',\vv',t+T)$ is the solution of the ordinary Lorentz-Boltzmann
equation for open systems $ f_S(\rv', \vv',t+T)$ with the initial
condition specified in Eq. (\ref{initial}). Due to time reversibility
and time translation invariance,
this probability is the same as
that for finding a particle at point $\rv$ with
velocity $-\vv$ at time t, that was at point $\rv'$
with velocity $-\vv'$ at time $t-T$. Thus,

\begin{equation}
\label{trev}
S(\rv,\vv,t|\rv',\vv',t+T) = S(\rv',-\vv',t-T|\rv,-\vv,t) =
S(\rv',-\vv',0|\rv,-\vv,T). 
\end{equation}

The integration in Eq. (\ref{survival}) with
respect to $\rv'$ and $\vv'$ allows us to replace $S(\rv,\vv,t|t+T)$
by the solution, $f_S(\rv,-\vv,T)
{\cal{N}} \delta(v-v_0)$,
of the Lorentz-Boltzmann equation with homogeneous initial condition
$f_S(\rv,-\vv,0) = 1$.
Here $\cal{N}$ is a normalization constant given by ${\cal{N}}=(2 \pi
 v_{0})^{-1}$ and $(4\pi v_0^{2})^{-1}$ in two and three dimensions,
respectively.
That is, we need the long time solution of the Lorentz-Boltzmann equation, with
absorbing boundary conditions, and with an initial condition which is
uniform in space and 
in velocity directions. Such a solution will not stay uniform
due to the escape of particles through the
absorbing boundaries.

Therefore Eq. (\ref{repelav}) is equivalent to
\begin{equation}\label{repelav2}
\langle g(\ro) \rangle_{Rep} = \lim_{t \to \infty}\lim_{T\to\infty}
\frac{\displaystyle 
\int d \ro 
d \rv d \vv F(\rv, \vv, \ro,t) f_S(\rv,-\vv,T) g(\ro)}{\displaystyle
\int d \ro d \rv d \vv F(\rv, \vv, \ro,t) f_S(\rv,-\vv,T)}
\end{equation}

Eq. (\ref{repelav2}) reduces the  calculation of  the sum of the
Lyapunov exponents and the maximum Lyapunov exponent on the repeller
to standard integrations, when the solution of the Lorentz-Boltzmann
equation for $f_S(\rv,-\vv,T)$ and that of the ELBE for $F$ are obtained. In the
following sections we will solve these
equations  for large systems with absorbing boundary conditions. For
long times, $t$, the
solution of the ELBE for $F$ will be obtained as a generalized
Chapman-Enskog \cite{chapcow} expansion of
the form 

\begin{equation}\label{ggradient}
F(\rv,\vv,\ro,t) = {\cal{N}}\delta(v - v_0)(\psi_0(\ro) n_m(\rv,t) +
\psi_1(\ro) \vv\cdot
\nabla n_m(\rv,t) + \psi_2(\ro) v^2 \nabla^2 n_m(r,t)+\cdots).
\end{equation}
 
Also $n_m(\rv,t)$ is the {\em slowest decaying eigenmode of the diffusion
equation} with the given absorbing boundary conditions,
\begin{equation}\label{boundary}
n_m(\partial V) = 0.
\end{equation}
In second order in the gradient we have kept only the scalar part $\nabla^2
n_m$. 
A possible contribution 
of order $\nabla^2$ to the Lyapunov exponents from a term
in Eq. (\ref{ggradient}) proportional  to the
traceless tensor ${\bf v} 
{\bf v} -( v^2/d) \, {\bf 1}$ vanishes after integration with respect to  the
velocity and is therefore neglected in this equation.
The solution of the usual Lorentz-Boltzmann equation for $f_S(\rv,-\vv,T)$ can
also be written as a Chapman-Enskog expansion 
\begin{equation}\label{gradient}
f_S(\rv,-\vv,T) = {\cal{N}}\delta(v - v_0) (n_m(\rv,T) - c_d
\vv\cdot \nabla n_m(\rv,T)+\cdots).
\end{equation} 
The constants in Eq. (\ref{gradient}) are $c_2 = -3/(4 \nu)$ and $c_3 = -1/\nu$ in two and three dimensions
respectively, where $\nu$ is the mean collision frequency. Due to the normalization
condition, Eq. (\ref{3.2}), the functions $\psi_i$ in Eq. (\ref{ggradient}) have to fulfill the
conditions 
\begin{eqnarray} 
\int d \ro' \psi_0(\ro') &=& 1, \label{psi0norm}\\  
\int d \ro' \psi_1(\ro') &=& c_d, \label{psi1norm}\\  
\int d \ro' \psi_2(\ro') &=& 0.  \label{psi2norm}   
\end{eqnarray}

It is important to note that in Eqs. (\ref{ggradient}) and
(\ref{gradient}), the eigenmodes $n_m(\rv,t)$ and $n_m(\rv,T)$ have
the functional forms $n_0(\rv)\exp[-t\omega]$ and
$n_0(\rv)\exp[-T\omega]$, respectively, where $\omega$ denotes the
eigenvalue of the slowest decaying mode of the diffusion equation, and
$n_0(\rv)$ is the corresponding eigenfunction.

Using Eqs. (\ref{ggradient} - \ref{psi2norm}) we
can write Eq. (\ref{repelav2}) as

\begin{equation} \label{repelav3}
\langle g(\ro) \rangle_{Rep} = \frac{\displaystyle \int d \ro \int d^d r \left[
  \psi_0 n_0^2(\rv) - v_0^2
(\frac{ c_d}{d} \psi_1 +  \psi_2) (\nabla
n_0(\rv))^2\right]g(\ro)}{\displaystyle \int
d^d r ( n_0^2(\rv) -
\frac{v_0^2 c_d^2}{d} (\nabla n_0(\rv))^2)} +\cdots.
\end{equation}
We note that exponentially decaying factors have canceled in the
numerator and denominator of Eq. (\ref{repelav3}).

Since we only kept terms up to the order $\nabla^2$ when deriving
(\ref{repelav3}), we also have to expand the 
denominator in Eq. (\ref{repelav3}). The final result for averaging a
quantity $g(\ro)$ on the
repeller, up to and including terms of order $\nabla^2$, is given by 

\begin{equation} \label{repelav4}
\langle g(\ro) \rangle_{Rep} =  \int d \ro \left\{ \psi_0  +
\left[\frac{c_d^2}{d} (\psi_0 - \frac{\psi_1}{c_d}) -  \psi_2\right]
v_0^2 \, \overline{q}^2 \right\}\; g(\ro) +\cdots,
\end{equation}

with 
\begin{equation}\label{q}
\overline{q}^2 = \frac{\displaystyle \int d^d r (\nabla 
n_0(\rv))^2}{\displaystyle \int
d^d r  n_0^2(\rv) }.
\end{equation}
The quantity $\overline{q}$  may be interpreted as a wave vector.
If, for example, a system is considered with absorbing boundaries at $x = \pm
L/2$, but of infinite extent in other directions, Eq. (\ref{q}) can be evaluated to show that  $\overline{q} \equiv q_x^{min} =
(\pi /L)$ is the smallest possible wave vector in the x direction. We
will consistently neglect effects due to microscopic, kinetic boundary
layers near $\partial V$, since such effects are unimportant in the
escape-rate formalism as the size of the system becomes large. 
 
That this is the case can be understood in the following way: The
fraction of the volume taken up by the boundary
layers is of order $\lambda/L$ with $\lambda$ the mean free path and
$L$ a  length of the order of the diameter of
the system. The density in the boundary layer is equally of order
$\lambda/L$  compared to the average density in
the system, as a result of the absorbing boundary condition. Finally
the  escape rate near the boundary is of
similar order, as it is proportional to the local density. As a result
the  effects of the boundary layer are of
order $(\lambda/L)^3$, whereas here we will be interested in terms of
order  $(\lambda/L)^2$ only.

\section{The Lyapunov exponents}

The strategy presented so far is applicable to the calculation of all
quantities which can be written as ensemble averages of functions of the ROC
matrix. In our case this always happens, if a quantity is given as a time
average over a function of $\ro(t)$. The sum of the positive Lyapunov exponents
fulfills
this property as shown by Sinai \cite{sinai}.  We {\em assume}
that trajectories on the repeller for the
Lorentz gas are 
sufficiently
ergodic, so 
that we  can write 
the sum of the exponents as an average over the appropriate ensemble
of ROC matrices (see also I). 
\begin{equation} \label{sinaiformel}
\sum_{\lambda_i >0}\lambda_i({\cal R}) = v \langle Trace(\ro^{-1})
\rangle_{Rep}. 
\end{equation} 
In two dimensions this is trivially also the largest exponent.  
The maximum Lyapunov exponent
in 3 dimensions is not 
calculated as a time
average, but as an average over a function of time of free flight and
collision parameters. 
The
separation of trajectories at a given time can then be written as product of
matrices, each describing the propagation of the ROC between two
collisions,
as explained in I \cite{vbld}. Using an identical method,
we find
\begin{eqnarray}\label{lmaxtime}
\delta \rv^\perp(t) &=& \prod_{i=1}^N \mbox{\boldmath
  $U$}_i(\tau_i,\phi_i,\alpha_i) 
\delta \rv^\perp(0), \;\;\; \mbox{with} \nonumber \\
 \mbox{\boldmath $U$}_i(\tau_i,\phi_i,\alpha_i) &=& T \exp(\int_{t_i}^{t_{i+1}}
 v\ro^{-1}(t',\phi_i,\alpha_i) dt'). 
\end{eqnarray}
Here $\phi_i$ and $\alpha_i$ are the scattering angles
at the $i$-th collision
and $\tau_i = t_{i+1} - t_i$ is the time of free flight between the $i$-th and
$i+1$-th collision. It is important to notice that
Eq. (\ref{lmaxtime}) cannot be written in terms of a
time integral over a local
function of $\ro(t)$, since the ${\bf{U}_i}$ matrices typically do not commute
with each other.

Since 
for dilute Lorentz gases correlations between collision events are not important in the
limit of low densities, 
we can make the
approximation that the matrices \mbox{\boldmath $U_i$} are
independent, randomly, and to the order in gradients that we use here,
isotropically distributed matrices, each with independent free flight 
times and collision parameters selected from appropriate
distributions, to be discussed below\cite{cpv}. We will postpone a
discussion of the isotropy of 
the ${\bf{U}}$ matrices until Section V and Appendix B, where we will calculate the
largest Lyapunov exponent for three dimensional systems. 
The maximum Lyapunov exponent is then given by 
\begin{equation}\label{lmaxgen}
\lambda_{max}({\cal R}) = \lim_{N \to \infty} \frac{1}{\sum_{i=1}^N
  \tau_i} \sum_{i=1}^N\ln (|\mbox{\boldmath$U_i$}\cdot \vec{e}|),
\end{equation}
where $\vec{e}$ is an arbitrary unit vector normal to the velocity $\vec{v}$. 
As in the case of infinite systems, we now 
write the right hand side of Eq. (\ref{lmaxgen}) as
an average 
over the distribution of the matrices \mbox{\boldmath$U_i$}. To do this we
have first to derive the distribution of particles, which collided
a time of free flight $\tau$ ago, 
$f_{F}(\rv,\vv,\tau,t)$, for an open 
system. This is done in Appendix A. Then we have to make
sure that the average is restricted to the repeller. This can be
achieved as described in Section II
by using the survival probability. We argue as follows: To determine
the appropriate average of $\ln (|\mbox{\boldmath$U_i$}\cdot
\vec{e}|)$, one needs the distribution of particles that have collided
at a point $\rv$, with the scattering angles $\phi,\alpha$, that have the
velocity $\vv$ after collision, and that travel freely for a time $\tau$
until the next collision. Further, to include only trajectories on the
repeller, we again need the survival probability, $S(\rv,\vv,t|t+T)$,
for a particle with 
$\rv,\vv$ at time $t$ to remain in the system at least until time
$t+T$. We can express the distribution of particles colliding at point
$\rv$ and having velocity $\vv$ after collision in terms of the
distribution of particles arriving to collide at point $\rv$ with the
{\em restituting} velocity $\vv\prime$, since the rate at which
particles arrive at a 
scatterer before collision
should be equal to the rate at which they leave the scatterer after
collision. The 
former rate can be calculated from kinetic theory without difficulty
due to the assumption of molecular chaos, while the latter rate
requires that correlations between particles and scatterers, produced
by a collision, be taken
into account. Note also that $|\vv\cdot\hat{n}| =
|\vv\prime\cdot\hat{n}|$, and that $\vv\cdot\hat{n} \geq 0$, while
$\vv\prime \cdot\hat{n}\leq 0$.
We finally arrive at the following
general expression 
for the maximum Lyapunov exponent of the open system,
\begin{eqnarray} \label{repellmax}
&&\lambda_{max}({\cal{R}})=\nu_{
cr} \langle \ln|\mbox{\boldmath$U$}\cdot
\vec{e}| \rangle_{cr}  \nonumber \\
&&= \lim_{
T \to \infty} \nu_{cr} \frac{\displaystyle\int d \tau d \hat{n}  
d \rv d \vv |\hat{n}\cdot\vv|\Theta(\hat{n}\cdot\vv) f_F(\rv, \vv\prime,
\tau,t) f_S(\rv,-\vv,
T) 
\ln |\mbox{\boldmath$U$}\cdot\vec{e}|}{\displaystyle\int d \tau d
\hat{n} d \rv d \vv |\hat{n}\cdot \vv| \Theta(\hat{n}\cdot\vv)
f_{F}(\rv, \vv\prime, \tau,t) 
f_{S}(\rv,-\vv,T)}. 
\end{eqnarray} 
Here
$<>_{cr}$ is used to distinguish the present average of variables that
are calculated only at the instants of collisions from the preceding
ones. Further the integrands in both numerator and denominator take into
account the effects of particles whose velocities are changed
from $\vv\prime$ to $\vv$ at time $t$, 
before
which a free flight time of
$\tau$ takes place. Thus, $f_{F}(\rv, \vv\prime, \tau,t)$ is the
density of particles with 
velocity 
$\vv\prime$, before collision, at the point $\rv$ and time $t$, which
had a time of free 
flight time of length
$\tau$ 
before the collision, and $\nu_{cr}$ is the average
collision frequency for trajectories on the repeller.  
Here 
\begin{equation}\label{vv'}
\vv\prime = \vv - 2 (\vv\cdot \hat{n}) \hat{n}.
\end{equation}
and $\hat{n}$
is the unit vector pointing from the center of the sphere (or disk) to the
point of impact. The factor $|\hat{n}\cdot \vv|$ appears in the
integrals when one 
takes into account the rate at which
collisions take place between the moving particle and the
scatterers. Eq. (\ref{repellmax}) should be compared with
Eq. (\ref{repelav}). The mathematical meanings of these averages are, of
course, different. In Eq. (\ref{repelav}) the function to be averaged
depends on the random matrix $\ro$ and the phase space variables
$\vec{r}$ and $\vec{v}$. In Eq. (\ref{repellmax}), the average 
depends on the random variables time of free flight $\tau$,
collision vector $\hat{n}$ and phase space variables $\vec{r}$ and
$\vec{v}$. But in both cases the average is restricted to the
repeller.

The distribution of times of free flight $f(\rv,\vv,\tau,t)$ and
therefore the average on the
repeller can be expressed as an expansion in gradients of the
density.
\begin{equation}\label{timegrad}
f_{F}(\rv,\vv,\tau,t) = \delta(v - v_0)(\psi^F_0(\tau) n_m(\rv,t) + \psi^F_1(\tau)
\vv\cdot\nabla n_m(\rv,t) + \psi^F_2(\tau) v^2 \nabla^2 n_m(r,t)).
\end{equation}
The functions $\psi_i^F$ are calculated in appendix \ref{A} for two and
three dimensions. 
By substituting Eqs. (\ref{timegrad}) and (\ref{gradient}) for $f_F$ and $f_S$, respectively, one finds that  through
second
order in the gradient, Eq. (\ref{repellmax}) assumes the form
\begin{equation}
\lambda_{max}({\cal R})=\nu_{cr} \frac{\int 
d\tau d\hat{n}
|\hat{n}\cdot\vec{v}|\Theta(\hat{n}\cdot\vec{v}) 
(\psi^F_0-(\frac{c_d}{d} \psi^F_1(1-2(\hat{n}\cdot\hat{v})^2) +
\psi_2^F)\bar{q}^2v_0^2) \ln |{\boldmath
U}\cdot\hat{e}|}{\int d\hat{n}  
|\hat{n}\cdot\hat{v}| 
\Theta(\hat{n}\cdot\hat{v}) (1-\frac{c_d^2}{d}
(1-2(\hat{n}\cdot\hat{v})^2)\bar{q}^2v_0^2)}  
\label{lyapmax2}.
\end{equation}
Here a number of simplifications have already been made. All terms that are odd in the velocity have been left out
since they
yield zero on integration. Second order terms in the denominator that give zero on integration
over the velocity due to the normalization of the Chapman-Enskog solution have been left out likewise. Furthermore
terms of the form $\hat{v}\hat{v}{\boldmath :}\nabla n\nabla n f(v)$ have been replaced by $\frac 1 d |\nabla n|^2
f(v)$, which gives the same result on integrating over velocity. Finally the factor $1-2\hat{v}\cdot\hat{n}$
results from integrating $\hat{v}'\cdot\nabla
n=(\hat{v}-2(\hat{v}\cdot\hat{n}) \hat{n})\cdot \nabla n$ over
$\hat{n}$. The components of $\hat{n}$ normal to $\hat{v}$ give
vanishing contributions. 
By defining the constant $A = \int d\hat{n} |\hat{n} \vec{v}| (1 - 2
(\hat{n}\cdot\hat{v})^2)/\int\,
d\hat{n}|\hat{n} \cdot \hat{v}|
\Theta(\hat{n}\cdot\hat{v})$, which  vanishes in $d=3$ and 
equals -1/3
in $d=2$, and expanding
Eq. (\ref{lyapmax2}) up to order
$\bar{q}^2$, we obtain the final result  
 
\begin{eqnarray} \label{repeltime}
\lambda_{max}({\cal{R}})&=& \nu_{cr} \int\,d\tau
d\hat{n}|\hat{n} \cdot \hat{v}|
\Theta(\hat{n}\cdot\hat{v})\left\{\psi^F_0 - \right. \nonumber \\
&&\left. \left ( \psi_2^F(\tau) -
\frac{c_d^2}{d} A \psi_0^F(\tau) + \frac{c_d}{d} 
\psi_1^F(1-2(\hat{n}\cdot\hat{v})^2) \right ) \bar{q}^2
v_0^2 \right\} \frac{\ln |{\bf U}\cdot \hat{e}|}{\int\,
d\hat{n}|\hat{n} \cdot \hat{v}|
\Theta(\hat{n}\cdot\hat{v})} .
\end{eqnarray}

\section{The Lyapunov exponent in 2 dimensions}

There is only one positive Lyapunov 
exponent in two dimensions. 
Therefore the positive Lyapunov exponent
can either be calculated with the help of Eq. (\ref{sinaiformel}),
i.e. formally
as the sum of the positive Lyapunov exponents or as the maximum
Lyapunov exponent with the help of Eq. (\ref{repellmax}). We will use both
methods, to demonstrate that the two approaches lead to the same
result. 
For the two dimensional Lorentz gas, the radius of curvature matrix
reduces to a scalar, $\rho$, the radius of curvature 
of two nearby trajectories (see \cite{vbd} and I), and the ELBE has
as variables $\rv,\vv$, and $\rho$. In this case, the ELBE,
Eq. (\ref{er5}) has the form \cite{vbd,vbld}

\begin{eqnarray}
 \lefteqn{\frac{\partial F}{\partial t}+\vec{v}\cdot\frac{\partial F}{\partial
     \vec{r}}+  v \frac{\partial}{\partial \rho} F + \nu F } \nonumber  \\
&= & 
 \frac{\nu}{2} \int_{-\pi/2}^{\pi/2}d\phi \cos\phi \,
\delta(\rho - \frac{a}{2} \cos \phi)
\times \label{3.1} \nonumber \\
&& f_B(\rv,\vv',t). 
\end{eqnarray}
Here we have supposed that the typical value of $\rho$ before
collision is of the order of the mean free path length, and the delta
function appearing in Eq. (\ref{er5}) has been simplified to that
appearing in Eq. (\ref{3.1}). The solution of the ELBE is normalized   
according to Eq. (\ref{3.2}).
With $\vv'$ we denote the velocity of a particle immediately before 
the collision, which results in a velocity $\vv$ after the scattering event
(\ref{vv'}). 
For the solution of Eq.
(\ref{3.1}) we impose the boundary condition that $F$ vanishes both at
zero and at infinite values of $\rho$, i.e.,
\begin{equation}
F(\rv,\vv,\rho = 0,t ) = \lim_{\rho \to \infty} F(\rv,\vv, \rho,t) = 0.
\end{equation}
We impose the boundary condition at $\rho=0$, because the radius of
curvature increases during free motion of the particle, collisions
never reduce its value to zero, and all trajectories with an initial 
negative value for the radius of curvature 
will acquire positive radius of curvature with probability one. The negative 
sign of the radius of curvature is conserved only on the 
stable manifold, which has Lebesgue measure zero in the phase space.      

The integration over $\phi$ can now be performed and we obtain from
Eq. (\ref{3.1}) 

\begin{eqnarray}
 \lefteqn{\frac{\partial F}{\partial t}+\vec{v}\cdot\frac{\partial F}{\partial
     \vec{r}}+  v \frac{\partial}{\partial \rho} F + \nu F } \nonumber  \\
&= & 
 \frac{\nu}{a} \Theta(1 -\sigma) \frac{\sigma}{(1-\sigma^2)^{1/2}} \,
\times \label{3.3}\\
&& (f_B(\rv,\vv'_+,t) + f_B(\rv,\vv'_-,t)). \nonumber
\end{eqnarray}

where $\sigma = 2 \rho/a \leq 1$. The velocities $\vv\prime_{\pm}$ result from the
evaluation of the delta function at precollisional
velocities which satisfy the relation $\vv\cdot \hat{n} =  \sigma
v$. We may use the  
expansion Eq. (\ref{gradient}) for the Lorentz-Boltzmann density
$f_B(\rv,\vv,t)$, by replacing $-\vv$ in Eq. (\ref{gradient}) by $\vv$. Then we use the
relation  
$\hat{n}(- \phi) = -\hat{n}(\phi)
+ 2 \cos \phi \; \hat{v} $  to obtain, from Eq. (\ref{3.3}), an equation
for the distribution function  
$\tilde{F}(\rv,\vv,\sigma,t) = a F(\rv,\vv,\rho,t)/2 $  given by  
\begin{eqnarray}
 \lefteqn{\frac{\partial \tilde{F}}{\partial t}+\vec{v}\cdot\frac{\partial
     \tilde{F}}{\partial 
     \vec{r}}+  \frac{2 v}{a} \frac{\partial}{\partial \sigma} \tilde{F} + \nu \tilde{F} }
     \nonumber   \\
&= & 
 \nu \Theta(1 -\sigma) \frac{\sigma}{(1-\sigma^2)^{1/2}} \,
\times \label{3.4} \nonumber \\
&& (n_m(\rv,t) - \frac{3}{4 \nu}  (1 - 2 \sigma^2) \vv\cdot\nabla
     n_m(\rv,t)).
\end{eqnarray}

Using the Chapman-Enskog {\it Ansatz}, Eq. (\ref{ggradient}), for $\tilde{F}$,
we obtain equations for
$\psi_0,\psi_1$ and $\psi_2$ by comparing equal orders of the gradients of the
density $n_m$ on both sides of Eq. (\ref{3.4}). The zeroth order
equation is (\cite{vbd} and I) 

\begin{equation}
\frac{2 v}{a} \frac{\partial}{\partial \sigma} \psi_0 + \nu \psi_0 
=   \nu \Theta(1 -\sigma) \frac{\sigma}{(1-\sigma^2)^{1/2}}, \label{3.5}
\end{equation}

with solution  

\begin{equation}\label{3.6}
\psi_0(\sigma) = \left \{ 
\begin{array}{lr} 
\frac{a \nu} {2 v} e^{-\nu \sigma a/(2v)} & \mbox{for}
   \; \sigma >1\\
\frac{a \nu}{2 v} [1-(1-\sigma^2)^{1/2}], & \mbox{for} \; \sigma <1
\end{array} \right.
\end{equation}
This solution is continuous at $\sigma=1$ and fulfills the normalization
condition Eq. (\ref{psi0norm})
up to corrections of relative order $\tilde{n} = na^{2}$. 

By equating the terms to first order in the gradient in the
density in Eq. (\ref{3.4}), we obtain an equation for $\psi_1$, given
by

\begin{equation} \psi_0(\sigma) +  \frac{2 v}{a} \frac{\partial}{\partial
    \sigma} \psi_1 + \nu \psi_1  
= 
 - \frac{3}{4}\Theta(1 -\sigma) \frac{\sigma (1-2
   \sigma^2)}{(1-\sigma^2)^{1/2}}  \label{3.7}
\end{equation}
Here we used the fact that the time derivative $\frac{\partial
n_m}{\partial t}$ is of second order in the gradient - via the
diffusion equation. 
The solution of Eq. (\ref{3.7}) satisfies the normalization  condition
Eq. (\ref{psi1norm}) with $c_2=- \frac{3}{4 \nu}$.
This can easily be seen by
integrating  both sides of Eq. (\ref{3.7}) with respect to $\sigma$.
The solution of Eq. (\ref{3.7}) has to be continuous at $\sigma =1$
and it is given by  

\begin{equation} \label{3.8}
\psi_1(\sigma) = \left\{ 
\begin{array}{lr} 
-\frac{\nu a^2}{4 v^2} \sigma e^{-\nu \sigma a /(2 v)} + \frac{a}{8 v}
   e^{-\nu \sigma a 
   /(2 v)} & \mbox{for}
   \; \sigma >1\\
\frac{a}{8 v} [1-(1-\sigma^2)^{1/2} (1 + 2 \sigma^2)], & \mbox{for} \; \sigma
   < 1.
\end{array} \right.
\end{equation}

By comparing the terms in order $\nabla^2 n_m$ in Eq. (\ref{3.4}), we
obtain the equation for $\psi_2$ given by

\begin{equation}\label{3.9}
\frac{D}{v^2} \psi_0(\sigma) + \frac{1}{2} \psi_1(\sigma) + \nu \psi_2(\sigma)
+ \frac
{2v} a \frac{d}{d \sigma} \psi_2(\sigma) =0 
\end{equation}

Here we used the diffusion equation for the density
$n_m$   
\begin{equation}\label{4.19}
\frac{\partial}{\partial t} n_m(\rv,t) = D \nabla^2 n_m(\rv,t),
\end{equation}
and the Chapman--Enskog solvability condition, Eq. (\ref{psi2norm}).
The low density value of the diffusion coefficient $D$ in two
dimensions is $D = (3 v^2)/(8 \nu)$. For our purposes it is sufficient
to write the solution of Eq. (\ref{3.9}) in the form
\begin{equation}\label{3.10}
\psi_2(\sigma) = - e^{- \frac{\nu a \sigma}{2 v}}\int_0^\sigma d \sigma'
  e^{\frac{\nu a \sigma'}{2 v}}  (\frac{D}{v^2} \psi_0(\sigma') + \frac{1}{2}
  \psi_1(\sigma')). 
\end{equation}

Once we know $\psi_0 ,\psi_1$ and $\psi_2$ we can obtain averages on the
repeller using Eq. (\ref{repelav4}). The positive Lyapunov exponent is
(see Eq. (\ref{sinaiformel})) 
\begin{eqnarray}
\lambda^+({\cal R}) &=& \frac{2 v_0}{a} \langle \frac{1}{\sigma} \rangle_{Rep}
\label{3.11} \\
\langle \frac{1}{\sigma} \rangle_{Rep} &=&  \int d \sigma ( \psi_0  +
(\frac{9}{32 \nu^2} (\psi_0 +\frac{4 \nu\psi_1}{3}) -  \psi_2)
v_0^2 \, \overline{q}^2 )\; \frac{1}{\sigma}\label{3.12}
\end{eqnarray}

Using Eqs. (\ref{3.6}, \ref{3.8}, \ref{3.10}) we can now calculate the
result in the 
lowest order in the density by standard integrations, and we obtain,
to second order in the gradients, 

\begin{equation}\label{3.13}
\lambda^+({\cal R}) = \lambda^+_0 + \left( \frac{\lambda^+_0}{\nu} -\frac{1}{2}\right) \; \; D \, \overline{q}^2. 
\end{equation}   
Here $\lambda^+_0$ is the equilibrium solution for the Lyapunov exponent of an
infinite system (\cite{vbd}, I) given by
\begin{equation}\label{3.14}
\lambda^+_0 =  2nav[-\ln(2na^2) +1 -{\cal C}],
\end{equation}
where ${\cal{C}}$ is Euler's constant.

We can also use the
method for the calculation of the largest Lyapunov exponent to obtain the
result for $\lambda^+$ as an average over the distribution of time of free
flight on the repeller.  
The rank of the matrix \mbox{\boldmath $U$}, defined in equation
(\ref{lmaxtime})  is unity in two 
dimensions, i.e $U$ is a scalar given by 
\begin{equation}\label{3.15}
U =  (1 + \frac{2 v\tau}{a\cos\phi}) 
\end{equation} 
The leading contribution in the low density approximation is obtained by
keeping only the term proportional to $\tau$ in Eq. (\ref{3.15}). The maximum 
Lyapunov exponent is then in leading order 
\begin{equation}\label{3.16}
\lambda^+({\cal R}) = \frac{1}{\langle \tau \rangle_{cr}} \langle \ln (\frac{2 v}{a} 
    \tau)  - 
\ln \cos\phi \rangle_{cr}.
\end{equation} 

By using Eqs. (\ref{timegrad}), (\ref{repeltime}), and (\ref{time2d}) for 
the distribution of times 
of free flight in two dimensions, we can easily recover the result,
Eq. (\ref{3.13}).  

\section{The Lyapunov exponents in 3 dimensions}

The ELBE for the three dimensional Lorentz gas can be solved by
choosing an appropriate parameterization of the 
(ROC) matrix
$\ro$. 
Starting from Eq. (\ref{er5}) we can simplify 
the delta function appearing in the restituting part of the collision 
operator (see I), so as to obtain

\begin{eqnarray}
 \lefteqn{\frac{\partial F}{\partial t}+\vec{v}\cdot\frac{\partial
F}{\partial \vec{r}}+ v\left(\frac{\partial}{\partial \rho_{11}}
+\frac{\partial}{\partial 
\rho_{22}}\right) F =} \nonumber \\ 
& &  -\nu F + 
 \frac{\nu}{\pi} \int_{0}^{\pi/2}d\phi\int_{0}^{2\pi}d\alpha\sin\phi\cos\phi
\prod_{i\le j}\delta(\rho_{ij}-\rho_{ij}(\phi,\alpha))
\times \label{4.1} \nonumber \\
&& \int d
\rho'_{11}\int d\rho'_{12}\int 
d\rho'_{22} F(\rv,\vv',\ro',t). 
\end{eqnarray}

Here the average (infinite system
and low density) collision frequency, $\nu$, is $\nu
=  n a^2 v \pi$.  The ROC matrix is symmetric, therefore only three parameters
are necessary to describe its behavior.
As shown in I it is convenient to use the
eigenvalues $\rho_1,\rho_2$ and the off diagonal element $\rho_{12}$ as
parameters. Together with the normalization condition Eq. (\ref{3.2}) 
and the solution of the Lorentz--Boltzmann equation as a gradient
expansion, given by Eq. (\ref{gradient}), with $-\vv$ replaced by
$\vv$, and $c_{3} = -1/\nu$,
\begin{equation} \label{4.3}
f_B(\rv, \vv,t) = {\cal{N}}\delta (v- v_0) (n_m(\rv,t) - \frac{1}{\nu}
\vv\cdot\nabla 
n_m(\rv,t)+\cdots),
\end{equation}
with $\cal{N}$ defined below Eq. \ref{initial}.
We find that Eq. (\ref{4.1}) can be written as 
\begin{eqnarray}
 \lefteqn{\frac{\partial\tilde{F}}{\partial t}+\vec{v}\cdot\frac{\partial
\tilde{F}}{\partial \vec{r}}+ v\left(\frac{\partial}{\partial \rho_{1}}
+\frac{\partial}{\partial 
\rho_{2}}\right){\tilde{F}}   + \nu\tilde{F} =} \nonumber \\ 
& &  
 \frac{\nu}{\pi} \int_{0}^{\pi/2}d\phi\int_{0}^{2\pi}d\alpha\sin\phi\cos\phi
\, |\cos 2 \alpha|\, \delta (\rho_1 - \frac{a}{2} \cos \phi) \delta(\rho_2
- \frac{a}{2 \cos \phi}) \times \label{4.4} \nonumber \\
&&\delta(\rho_{12} + ( a \cos \phi \; \tan^2 \phi \;
\sin 2\alpha)/4) \; f_B(\rv,\vv',t).
\end{eqnarray}
The factor $|\cos2\alpha|$ results from the transformation from the
variables $\rho_{11},\rho_{22},\rho_{12}$ to
$\rho_1,\rho_2,\rho_{12}$ in the $\delta$-functions in the right hand side of
Eq. (\ref{4.1}), and we set 
$\tilde{F}(\rho_1,\rho_2,\rho_{12})=F(\rho_{11},\rho_{22},\rho_{12})$.
Please note, that $\tilde{F}$ is not yet normalized with respect to
$\rho_1, \rho_2, \rho_{12}$.  
The velocity $\vv\prime$ before the collision depends on the collision
angles $\alpha$ and $\phi$, through the relation
\begin{eqnarray}
\vv\prime & = &\vv-2(\hat{n}\cdot\vv)\hat{n}, \; \; {\mbox{with}}\nonumber \\
\hat{n}   & = & \cos\phi\,\hat{v}+\sin\phi\cos\alpha\,\hat{v}_{\perp,1}+
\sin\phi\sin\alpha\,\hat{v}_{\perp,2},
\label{3vex}
\end{eqnarray}  
and the unit vectors $\hat{v},\hat{v}_{\perp,1},\hat{v}_{\perp,2}$
form an ortho-normal set.

Now, using Eq. (\ref{4.3}), we can perform the $\alpha$ integration,
and we keep only the non-vanishing terms in the $\phi$
integration. After integrating over $\alpha$ and making $f_B$
explicit, we find that we can rewrite Eq
. (\ref{4.4}) as
\begin{eqnarray}
 \lefteqn{\left[\frac{\partial\tilde{F}}{\partial t}+\vec{v}\cdot\frac{\partial
\tilde{F}}{\partial \vec{r}}+ v \left(\frac{\partial}{\partial \rho_{1}}
+\frac{\partial}{\partial 
\rho_{2}}\right)\tilde{F}   + \nu \tilde{F}\right]} \nonumber \\ 
& = &   \frac{8 \nu{\cal{N}}}{\pi a} \int_{0}^{\pi/2}d\phi \; \sin\phi\cot^{2}
\phi
\,\Theta\left(1-\left|\frac{4\rho_{12}}{a\cos\phi\tan^{2}\phi}\right|\right)
\delta(\rho_1
-\frac{a\cos\phi}{2})\times
\nonumber \\
& & \delta(\rho_2 - \frac{a}{2\cos\phi})
\delta (v- v_0) \left(n_m(\rv,t) - \frac{1}{\nu} ( 1 - 2
\cos^{2}\phi) \vv\cdot\nabla
n_m(\rv,t)+\cdots\right). 
\label{4.6}
\end{eqnarray}
Here we have explicitly indicated only the terms needed for our
further calculations. It is now convenient to introduce the definitions $2 \rho_i/a = \sigma_i$ and
$2 \rho_{12}/a = 
\sigma_{12} $, and to perform the $\phi$ integration. We find that
\begin{eqnarray}
 \lefteqn{\left[\frac{\partial\tilde{F}}{\partial t}+\vec{v}\cdot\frac{\partial
\tilde{F}}{\partial \vec{r}}+ 2 v/a \left(\frac{\partial}{\partial \sigma_{1}}
+\frac{\partial}{\partial 
\sigma_{2}}\right) \tilde{F}   + \nu \tilde{F}\right] } \nonumber \\ 
&=&  \frac{32 \nu{\cal{N}}}{\pi a^3} \Theta\left(1-\left|\frac{2\sigma_1\sigma_{12}}{1-\sigma_{1}^2}\right|\right)
\Theta(1-\sigma_1) \frac{\sigma_1^2}{1-\sigma_1^2}
\,  \delta(\sigma_2
- \frac{1}{\sigma_1}) \times  \nonumber \\
&& \delta (v- v_0)
\left(n_m(\rv,t) - \frac{1}{\nu} ( 1 - 2 \sigma_1^2) \vv\cdot\nabla
n_m(\rv,t)+\cdots\right). 
\label{4.7}
\end{eqnarray}

We point out that $\tilde{F}$ is not normalized with respect to $\sigma_1, 
\sigma_2$ and 
$\sigma_{12}$, since we have not yet introduced the appropriate Jacobian.
 Equation (\ref{4.7}) can be further simplified by introducing a new set of
variables ($0 \le s , \, 0 \le z \le 1, \, -\pi/2 \le \gamma \le \pi/2$),
defined by the relations

\begin{eqnarray} \label{4.10}
&&\sigma_1 = z + s, \nonumber \\ 
&& \sigma_2 =\frac{1}{z} +s\\
&&\sigma_{12} = \sin
(\gamma)
\frac{1 -z^2}{2 z} \nonumber. 
\end{eqnarray}
 
The {\em normalized} distribution function $\tilde{f}
(\vec{r},\vec{v},s,z,\gamma,t) = J(s,z,\gamma)
\tilde{F}(
\vec{r},\vec{v},\rho_1,\rho_2,\rho_{12},t)$ obeys the equation 

\begin{eqnarray}
 \lefteqn{\frac{\partial \tilde{f}}{\partial t}+ \tilde{\partial} \tilde{f}  +
\frac{2v}{a} \frac{\partial}{\partial s} \tilde{f}   + \nu \tilde{f} }
\nonumber \\  
&=&  \frac{2\nu{\cal{N}}}{\pi} z \, \Theta(1 -
z)\Theta(z)\Theta(\pi/2 -\gamma )  \Theta(\gamma + \pi/2) 
\delta(s-0^{+}) \label{bneq}  \\
&& \delta (v- v_0)
\left(n_m(\rv,t) - \frac{1}{\nu} ( 1 - 2 z^2) \tilde{\partial}
n_m(\rv,t)\right), \nonumber
\end{eqnarray}
where $\tilde{\partial} = \vv\cdot \nabla$, and
$J(s,z,\gamma)$ is the Jacobian given by
\begin{eqnarray} \label{4.8}
J(z,s,\gamma) &=&
|\frac{\partial(\rho_{11},\rho_{22},\rho_{12})}{\partial(\rho_{1},\rho_{2},
\rho_{12})}|\frac{\partial(\rho_{1},\rho_{2},
\rho_{12})}{\partial(\sigma_1, 
\sigma_2,\sigma_{12})}| |\frac{\partial(\sigma_1,
\sigma_2,\sigma_{12})}{\partial(s,z,\gamma)}| \\
&=& \frac{a^3}{16} \frac{(1+z^2)(1-z^2)}{z^3}  .
\end{eqnarray} 
We point out that this Jacobian is independent of $s$. It guarantees the 
proper normalization of $\tilde{f}$ as

\begin{equation}\label{norm3}
\int_0^\infty d s \int_0^1 d z \int_{-\pi/2}^{\pi/2} d
\gamma \tilde{f}(\rv,\vv,s,z,\gamma,t) = f_B(\rv,\vv,t).
\end{equation}
where $f_B$ is the solution of the standard nonequilibrium Lorentz-Boltzmann
equation.

The physical meanings of $s$, $z$ and $\gamma$ become a bit more
transparent if we consider Eq. (\ref{4.10}) and Eq. (\ref{bneq}) in more
detail. We begin by noting, that the  
distribution function, $F(\rv,\vv,
\rho_1,\rho_2,\rho_{12},t)$, for the two eigenvalues and the off
diagonal element of the ROC matrix is established
through the dynamic process involving intervals of free flight
separated by collisions of the moving 
particles with the scatterers. The time dependence of the ROC matrix
can be completely 
expressed in terms of scattering angles and time of free flight. In
(I) it was shown that the eigenvalues of the ROC matrix increase
linearly in time during a free flight and the off diagonal element
stays constant.
The reparameterization Eq. (\ref{4.10})
exactly reflects this behavior. We can see that the dimensionless parameter s
corresponds to the 
time of flight, which is also clear since the $\delta$ - function
in $s$ in the  right hand side of 
Eq. (\ref{bneq}) shows that the gain term is a source of particles
with a free flight time of zero. From
Eq. (\ref{4.10}) we can then conclude that $z$, respectively
$\frac{1}{z}$, correspond to the 
possible values of normalized eigenvalues $\sigma_1$, $\sigma_2$ at
the collision. Equation (\ref{4.4}) shows that $z$ is physically the
cosine of the scattering angle $\phi$. With the same arguments, it can be seen
that
up to a factor of $2$, $\gamma$ finds it physical correspondence in the azimuthal 
scattering angle $\alpha$.   

As one might suspect here, the solution of Eq. (\ref{bneq}) can be interpreted 
as a joint distribution function for time of free flight and collision parameter.
This may seem surprising at first, because the latter are statistically
independent 
quantities. Correlations in $\tilde{f}$ however, are the result of considering this 
function at fixed time and position. In a comoving frame, i.e.\ cconsidering
$\tilde{f}(\vec{r}-\frac a 2 \sigma \hat{v},\vec{v},s,z,\gamma,t-a\sigma/2v)$,
one would find the variables $s$, $z$ and $\gamma$ to be uncorrelated indeed.

The
distribution of times of free flight in 3 dimensions Eq. (\ref{a1})
can 
be recovered, if we integrate Eq. (\ref{bneq}) over $z$ and
$\gamma$ and identify $a s/(2v)$ with $\tau$.

\begin{equation} \label{4.12}
f_F(
\vec{r},\vec{v},\tau)=
\frac a {2v} \int_{-\pi/2}^{\pi/2} d \gamma \int_0^1 dz
\tilde{f}(\vec{r},\vec{v},
\frac{2 v}{a} \tau,z,\gamma)
\end{equation}

For large systems, we can solve Eq. (\ref{bneq}), as in the 2
dimensional case, by using a gradient expansion 
\begin{eqnarray}\label{4.13}
\tilde{f} = \delta(v - v_0)
{\cal{N}} \Theta(\pi/2 -\gamma)
\Theta(\gamma +\pi/2 ) (n_m(\rv,t)
\, \tilde{\psi}_0(s,z) + \frac{a}{ 2v} \tpart n_m(\rv,t) \,
\tilde{\psi}_1(s,z) \nonumber \\ + 
\frac{a^2}{4}\nabla^2 n_m(\rv,t)
\,\tilde{\psi}_2(s,z)+\cdots).    
\end{eqnarray}
The solution of 
 Eq.\ (\ref{bneq}) has to satisfy the relation (\ref{norm3})
The quantity $\tilde{\psi}_0(s,z)$ was already obtained in I. It satisfies the equation
\begin{equation}\label{4.14}
\frac{d}{ds} \tilde{\psi}_0   + \tilde{\nu} \tilde{\psi}_0 = 
\nonumber \frac{
2\tilde{\nu}}{\pi} \,
\, \Theta(1-z)
\Theta(z) z \, \delta(s-0^+), 
\end{equation}
where the dimensionless collision frequency is $\tilde{\nu} = \frac{a}{2 v} 
\nu$. The solution of Eq. (\ref{4.14})
 is  
\begin{equation}\label{4.15}
\tilde{\psi}_0(s,z) = \frac{
2\tilde{\nu}}{\pi} \,
\, \Theta(1-z)
\Theta(z) z \, \Theta(s) e^{- \tilde{\nu} s}.
\end{equation}
The equation for $\tilde{\psi}_1$ is obtained by keeping only the terms
proportional to $\tpart n_m$
\begin{equation}\label{4.16}
\tilde{\psi}_0 + \frac{d}{ds} \tilde{\psi}_1   + \tilde{\nu} \tilde{\psi}_1 = 
\nonumber \frac{-
2
}{\pi} \,
\, \Theta(1-z)
\Theta(z) z (1 - 2 z^2)\, \delta(s
-0^+). 
\end{equation}
This equation is easily solved to give
\begin{equation}\label{4.17}
\tilde{\psi}_1(s,z) = \frac{-
2
\tilde{\nu}}{\pi} \, (s z + \frac{1}{\tilde{\nu}}
z (1-2 z^2))  
\, \Theta(1-z) \,
\Theta(z) \, \Theta(s) e^{- \tilde{\nu} s}.
\end{equation}
Considering the terms of order $\nabla^2 n_m(\rv,t)$ and keeping only
the scalar part of
$\tpart \, \tpart$, we obtain the equation
for $\tilde{\psi}_2$
\begin{equation} \label{4.18}
\hat{D} \tilde{\psi}_0 + \frac{1}{3}
 \tilde{\psi}_1 + \frac{d}{d s}
\tilde{\psi}_2 + \tilde{\nu} \tilde{\psi}_2 = 0. 
\end{equation}
Here the Chapman Enskog solvability condition Eq. (\ref{psi2norm}) 
and the diffusion equation for $n_m$, Eq. (\ref{4.19}) 
 \,
were used. We also introduced the dimensionless diffusion coefficient 
$\hat{D} = \frac{2}{a v} D $.  
The solution of this equation is given by 
\begin{equation}\label{4.20}
\tilde{\psi}_2(s,z) = \frac{
2\tilde{\nu}}{\pi} \, (-\hat{D} s z + \frac{1}{6} s^2 z + 
\frac{1}{3 \tilde{\nu}} s z (1-2 z^2)) 
\, \Theta(1-z) \,
\Theta(z) \, \Theta(s) e^{- \tilde{\nu} s}.
\end{equation}

The formulas for the sum of the Lyapunov exponents and the maximum
Lyapunov exponent
in equilibrium were obtained 
in I. The same 
formulae
are valid in the nonequilibrium case, if the averages are replaced by
averages on the repeller.
For the sum of the Lyapunov exponents  Eq.\ (\ref{repelav4})
gives

\begin{eqnarray} \label{4.21}
\frac{a}{2 v} \sum_{\lambda_i >0}\lambda_i({\cal R}) &=&
\langle(\frac{1}{z + s} + \frac{1}{1/z+s}) \rangle_{Rep},\\
\end{eqnarray}
while Eq. (\ref{repeltime}) leads to
\begin{eqnarray}
\lambda_{max}({\cal R}) &=& \nu_{cr}
\langle \ln |{\bf U}(\cos \phi,\alpha,\tau)\cdot 
\vec{e}_\psi| \rangle_{cr}, \label{4.21a}
\end{eqnarray}
with the average collision frequency on the repeller $\nu_{cr} =
\frac{1}{\langle \tau \rangle_{cr} }$ and

\begin{equation}\label{4.22}
|{\bf U}(z,\alpha,\tau)\cdot \vec{e}_\psi| =  
\frac {2v \tau}{a} \sqrt{\frac{1+z^4 + (1-z^4)\cos 2(\psi -\alpha)}{2
    z^2}}
\end{equation}
with $z= \cos \phi$,
as derived in I. 
The angle $\psi$ specifies the direction of the unit vector $\vec{e}_\psi$
in the  plane perpendicular to the trajectory.
In general an additional average over 
its stationary distribution $P(\psi)$ 
is
necessary. However, it can easily be shown that  
the corrections to an isotropic 
distribution of directions $P(\psi)$ are   
at most of order 
$\cos(\psi) \nabla^2 n(r,t)$  and do not contribute to the average 
in Eq. (\ref{4.21a}) in order $\nabla^2 n(r,t)$ (see appendix B). 

Now, using the equations Eqs. (\ref{4.21},  \ref{repelav4}), we are
led to the determination 
of the sum of the positive Lyapunov exponents, and we obtain, to
second order in the gradients,  

\begin{equation}\label{4.23}
\lambda^{+}_{max}({\cal R})+\lambda^{+}_{min}({\cal R}) = h_{KS}^0  -
D  \overline{q}^2   \; ( 
1+ 2 (\ln (\tilde{n} /2)  \, + {\cal C })),
\end{equation}
with $\tilde{n} =n \pi a^3
=2\tilde{\nu}$. Here, 
$h_{KS}^0$ is the KS-entropy for
an infinite system [I] at equilibrium given by
(\ref{hks0}), ${\cal C}$ is Euler's constant, and $\overline{q}$ is
defined by   
(\ref{q}), and     

\begin{equation}\label{hks0}
h_{KS}^0 = 2na^{2}v\pi[-\ln(\tilde{n}/2) -{\cal
  C}]+\cdots .
\end{equation} 
with the dots indicating higher density corrections.

This expression for the sum of the Lyapunov exponents can also be
calculated by averaging  
$ \ln \det \mbox{\boldmath $U$} = \mbox{Trace} \ln \mbox{\boldmath $U$}$
over the  
distribution of times of free
flight and the scattering angles $\Omega$ with the help of
Eq. (\ref{repeltime}), and by replacing $\ln|\mbox{\boldmath$U$} \cdot
\vec{e}|$ with $\mbox{Trace} \ln \mbox{\boldmath $U$}$
In leading order for large times of free flight i.e. small 
density of scatterers, we obtain 
\begin{equation}\label{4.24}
\lambda^{+}_{max}({\cal R})+\lambda^{+}_{min}({\cal R}) = \nu_{cr}
\langle \mbox{Trace} \ln (v \ro_+^{-1} \tau) \rangle_{cr}. 
\end{equation}
The matrix $\ro_+$ is the ROC immediately after a scattering event. It depends
in leading order in the density only on the scattering angles $\phi$ and
$\alpha$ and is defined in paper I. Here it is only important to notice that
it has eigenvalues $\rho_1 = \frac{a}{2} \cos \phi$ and $\rho_2 = \frac{a}{2
  \cos \phi}$. The trace of the logarithm in Eq. (\ref{4.24}) leads
therefore to a 
cancellation 
of the terms depending on $\ln \cos \phi$, so that we only have
to evaluate 
\begin{equation}\label{4.25}
\lambda^{+}_{max}({\cal R})+\lambda^{+}_{min}({\cal R}) = 2 \nu_{cr}
 \langle  \ln (\frac{2 v}{a}  
 \tau) \rangle_{cr}. 
\end{equation}
This strategy also leads to the result Eq. (\ref{4.23}).

With the help of Eqs. (\ref{4.21a}, \ref{4.22}, \ref{repellmax})
and Appendix A the
maximum Lyapunov exponent is given, to second order in the gradients, by 
\begin{equation}\label{4.26}
\lambda^{+}_{max}({\cal R}) = \lambda_{max}^0  + D \overline{q}^2 (
-\ln (\tilde{n}/2) - 
{\cal C} + \frac{1}{4} - \ln 2 ) \; ,
\end{equation}
where $\lambda_{max}^0$ is the equilibrium value of the maximum Lyapunov 
exponent  for an infinite system 
\begin{equation} \label{lmax0}
\lambda^{0}_{max} = na^{2}v\pi[-\ln(\tilde{n}/2) +\ln 2 -\frac{1}{2}-{\cal
C}]+\cdots. 
\end{equation}

The expression for the smallest positive Lyapunov exponent can be
obtained, to second order, 
from Eqs. (\ref{4.23}) and (\ref{4.26}), as

\begin{equation}\label{4.27}
\lambda^{+}_{min}({\cal R}) = \lambda_{min}^0  + D \overline{q}^2 ( -\ln (\tilde{n}/2) -
{\cal C} - \frac{5}{4} + \ln 2 ) \cdots \; ,
\end{equation}
 
with  
\begin{equation} \label{lmin0}
\lambda^{0}_{min} = na^{2}v\pi[-\ln(\tilde{n}/2) -\ln 2 + \frac{1}{2}-{\cal
C}]+\cdots. 
\end{equation}

\section{Discussion}

We have now calculated, analytically, the spectrum of
positive Lyapunov 
exponents on the repeller for the open, dilute, random Lorentz gas in
two and three 
dimensions. Then using the escape-rate formula we may infer the values of the KS entropies on the repeller as well. We find
that the corrections to the equilibrium values of the Lyapunov
exponents and KS entropies are of order $1/L^{2}$ where $L$ is some
characteristic size of the open system. We should point out that
Gaspard has discussed a reformulation of the escape-rate formula so as
to be able to express the diffusion coefficient in terms of the
Hausdorff dimension of the fractal repeller \cite{gasbook}. Using his
method we can easily see that the dimension of the fractal repeller is
slightly less than the embedding dimension ($3$ for $d=2$, and $5$ for
$d=3$) by terms of order $1/L^{2}$. These results are to be expected
for the Lorentz gas, since diffusion is normal, and the
escape-rate formula should be free of difficulties.

Gaspard and Baras \cite{gaspbaras} have examined the chaotic properties of
the periodic open Lorentz gas at sufficiently high densities that
there are no infinite horizons for the moving particle. They used
numerical simulations, and obtained independent results for the
Lyapunov exponents and for the KS entropy on the repeller, as functions of
the system
size, $L$. Then they compared these results
with numerical and with approximate analytical values for the
diffusion coefficient and found good agreement. 

At the present time there are no computer simulations of open, random
Lorentz gases, to which our analytic results can be
compared, but we may compare our results to intuitive expectations. To do this we will need to
note that the average collision frequency, $\nu_{cr}$ or
equivalently, the mean free 
time between collisions on the repeller, $\tau_{cr}=1/\nu_{cr}$,
differ from the corresponding 
quantities in an infinite system. In fact $\tau_{cr}$ can explicitely
be calculated by using an expression
analogous to Eq. (\ref{repeltime}), where $\ln |{\bf U} \cdot \hat
{e}|$ is replaced by $\tau$ and the factor $\nu_{cr}$ is
dropped. Together with the 
expressions for $\tilde{\psi}_i^F$ in two and three dimension
Eqs. (\ref{timegrad}, \ref{time2d}, and 
\ref{time3d}) we obtain an expression for the mean free time between
collisions on the repeller given by 

\begin{equation} \label{srate}
\tau_{cr} \equiv \frac{1}{\nu_{cr}} =  \frac{1}{\nu} - \frac{D}{\nu^2}
\overline{q}^2.  
\end{equation} 
  
This result shows that the mean free time on the repeller is smaller 
than that in the infinite
system. In addition, its dependence on the escape-rate and equilibrium
collision frequency is the same in two and three dimensions. Thus, we
see that trajectories on the repeller are constrained to have higher
collision frequency than those for the infinite, equilibrium system.

The Lyapunov exponents i.e. the rates
of separation of trajectories, are 
larger on the repeller than in the infinite system. One might
have expected, that due to the restriction of the available phase space
volume for a particle on the repeller, the rate of separation might have
been smaller on the repeller, but the increased scattering rate 
counteracts this effect. However, if we compensate for this effect by
expressing the Lyapunov exponents in 
units of the mean free time on the repeller, the terms proportional to
$\overline{q}^2$ lead to a decrease of the Lyapunov exponents
because on average $\ln (\lambda/a)$ decreases.  Using
Eqs. (\ref{3.13}) and (\ref{srate}) for two dimensions and Eqs. (\ref{4.23},
 \ref{4.26}, \ref{4.27}) and (\ref{srate}) for three dimensions, the
Lyapunov exponents in natural units on the repeller are given by 
\begin{equation}
\lambda^+({\cal R})\, \tau_{cr} = \frac{\lambda^+_0}{\nu} -
 \frac{1}{2} \frac{D}{\nu} 
 \, \overline{q}^2 \; \; \mbox{ in 2 dimensions},
\end{equation}
and for three dimensions
\begin{eqnarray}
\lambda^{+}_{max}({\cal R}) \, \tau_{cr} &=& \frac{\lambda_{max}^0}{\nu}  -
\frac{D}{\nu} 
\overline{q}^2  \left(\frac{8 \ln 2 - 3}{4}\right) \, ,\nonumber \\
\lambda^{+}_{min}({\cal R}) \, \tau_{cr} &=& \lambda_{min}^0  - \frac{D}{\nu}
\overline{q}^2  \left(\frac{7 - 8 \ln 2}{4}\right), \nonumber \\
(\lambda^{+}_{max}({\cal R})+\lambda^{+}_{min}({\cal R})) \,
\tau_{cr} &=& \frac{h_{KS}^0}{\nu}  - 
\frac{D}{\nu} \overline{q}^2. 
\end{eqnarray}

It is also interesting to compare the KS entropy on the repeller with
its value in an infinite system. To make this a bit more concrete, we
consider a slab geometry, i.e. absorbing walls at $x=\pm 
L/2$, and
\begin{equation}
\overline{q} = \frac{\pi}{L}. 
\end{equation}
Then with Eqs. (\ref{3.13}) and (\ref{4.23}) respectively, we obtain.
\begin{eqnarray}
h_{KS}({\cal R}) &=& h_{KS}^0 + ( \frac{
h_{KS}^0}{\nu}
-\frac{3}{2}) \; \; D \,  
(\pi/L)^2 \, \; \mbox{in two dimensions} \\
h_{KS}({\cal R}) &=& h_{KS}^0  +    (
\frac{h_{KS}^0}{\nu} -2) \; D  (\pi/L)^2 \; \mbox{in three dimensions}
\end{eqnarray}
Thus the KS entropy increases above its infinite system value when
measured in standard time units. As in the case of the
Lyapunov exponent, this trend is only due to the increased scattering
rate on the repeller. When measured in natural units on the repeller,
the KS entropy has an especially simple form.

\begin{eqnarray}
h_{KS}({\cal R}) \; \tau_{cr} &=& \frac{h_{KS}^0}{\nu} - \frac{3}{2} \; \;
\frac{D}{\nu} \,  
(\pi/L)^2 \, \; \mbox{in two dimensions} \\
h_{KS}({\cal R}) \; \tau_{cr} &=& \frac{h_{KS}^0}{\nu} - 2 \; \;
\frac{D}{\nu} \,  
(\pi/L)^2 \,  \; \mbox{in three dimensions}
\end{eqnarray}

We conclude with a number of points:
\begin{enumerate}
\item The principal problem with the escape-rate method as an
analytical method for computing
transport coefficients, apart from the inherent difficulties involved
in calculating dynamical quantities, is that as yet we have no
analytical methods
for calculating the KS entropy on the repeller, independently of the
escape-rate formula. At the moment we can only use analytic techniques
to calculate the diffusion coefficients and the Lyapunov exponents,
leaving the KS entropy as a quantity to be derived from them.
\item It would be valuable to have some results from computer
simulations with which to compare the results obtained here.
\item As mentioned earlier, the information
and  Hausdorff dimension
of the fractal  repellers in both two and
three dimensions are very close to the full phase space dimensions,
three, for the Lorentz gas on the plane, and five, for the Lorentz gas
in space, but are smaller than these values by terms of order
$L^{-2}$. This is a simple consequence of the application of
Kaplan-Yorke type formulae to fractal repellers
\cite{huntpre}. For the information dimension we obtain 
\begin{eqnarray}
d_I &=& 3 - 2 \frac{\gamma}{\lambda^{+}_0} + O(1/L^4) \; \; \mbox{for} \; d =
2 \label{dinf2}\\ 
d_I &=& 5 - 2 \frac{\gamma}{\lambda^0_{min}} + O(1/L^4) \; \;
\mbox{for}\; d = 3 \label{dinf3} 
\end{eqnarray}
where $\gamma$ is the escape rate and $\lambda^{+}_0$, $\lambda^0_{min}$
are the positive Lyapunov exponent, and the smaller of the positive Lyapunov exponents in the
infinite system, respectively. 
Gaspard and Baras have used these fractal  dimensions to
express the diffusion coefficient in terms of the Hausdorff and
information dimensions of the fractal repeller for the two dimensional
case \cite{gasbook,gaspbaras}. 
\item A problem for further study is to extend the calculations given
here to a system
of many interacting particles such as gases of hard
disks or hard spheres. Some progress in this direction has been made,
and it is now possible to get analytic results for the KS entropy and
the largest Lyapunov exponents for dilute hard disk or hard sphere gases, in
equilibrium, in the thermodynamic limit \cite{vanzon,ks1,ks2}. It
would be very interesting to apply the escape-rate formalism to the
transport coefficients, such as the shear and bulk viscosities, and
thermal conductivity, appropriate for fluid systems, using the method
of Ref. \cite{dorgas}, and to determine the effects of the fractal
repeller on the dynamical quantities.  
\item The thermodynamic formalism for hyperbolic chaotic systems
provides a very useful method for expressing 
many of the chaotic
properties of 
both open 
and closed systems in terms of one
quantity, the topological pressure\cite{gasdo}. The use of kinetic
theory to evaluate the topological pressure for a
dilute random Lorentz gas
should certainly be possible, but has not yet been 
undertaken.

\item In the next paper in this series we
will consider the case of a
Lorentz gas with a charged moving particle placed in an external electric field
as well as in a random array of scatterers. Then a Gaussian thermostat
is applied which keeps the kinetic energy of the moving particle
fixed. 
The system eventually reaches a nonequilibrium steady state.
 We will calculate dynamical properties of the moving particle in
this nonequilibrium steady state and compare with the results of
computer simulations. A preliminary version of this work has already been
published \cite{vbdcpd,lvbd}.
\end{enumerate}

\section{Acknowledgments}
A.L. was partially supported by the SFB 262 of the Deutsche
Forschungsgemeinschaft.  
JRD wishes to acknowledge support from the
National Science Foundation under Grant NSF PHY-96-00428.
HvB acknowledges support by FOM
and by the NWO Priority Program Non-Linear
Systems, which are financially supported by the "Nederlandse Organisatie voor 
Wetenschappelijk Onderzoek (NWO)"

\begin{appendix}  

\section{The distribution of times of free flight } \label{A}

It is useful to have an expression for the mean collision frequency or,
equivalently,for
the mean free time between collisions for trajectories on the
repeller. Here we argue that such 
expressions can be obtained in a
simple way from a Lorentz-Boltzmann equation, modified so as to include a new
variable
$\tau$ which is the time since the last collision. 
For the distribution of particles
surviving in the system at time $t$,
with position 
$\rv$, velocity $\vv$, and with time $\tau$ since the last collision
we propose the equation
\begin{eqnarray}\label{a1}
\frac{\partial}{\partial t}f_F + \vv\cdot \nabla f_F + 
\dot{\vv} \cdot \frac{\partial}{\partial \vv} f_F + 
  \frac{\partial}{\partial \tau} f_F = - \nu f_F  \nonumber \\
\end{eqnarray}
with
\begin{equation} \label{a2}
f_F(\rv,\vv',\tau=0, t)= n a^{d-1} v  \int  d \hat{n}\Theta(-\hat{n}\cdot\vv)
|\hat{n}\cdot \hat{v}| 
  f_B(\rv,\vv', t).
\end{equation}
Eq. (\ref{a1}) can be derived in a way similar to the heuristic
derivation of the usual Lorentz-Boltzmann equation, but a few changes have to
be 
made.
Scattering and absorption at the 
boundary are the
only mechanisms which will reduce the number of particles with a  certain time
of free flight. Scattering will also act as a source of particles but
always with a time of free flight $\tau = 0$. This effect is taken
care of by the initial condition Eq. (\ref{a2}). 
The dependence of the distribution of times of 
free flight on 
both position and velocity is due to 
the non-uniformity in these variables of $f_B$ resulting from the
absorbing boundary  
condition in combination with a higher survival rate for particles
that collide more frequently 
(they diffuse more slowly). We note one important difference, which
will become crucial, if   the effect on external fields is considered. 
In the Boltzmann equation  for the one particle phase space density
$f_B$,  
which
is based on the consideration of the number of particles in a fixed
volume element in phase space, the streaming term is derived from
$\frac{\partial}{\partial q}(\dot{q} f_B(q,t)) = 0$ with
$q=(\vec{r},\vec{v})$. This is due to the   
conservation of particles in the absence of scattering events.  In the 
derivation of $f_F$ we have to consider the number of particles with a 
certain times of free flight $\tau$ after the last scattering event
and count how many of them are still there a time step $d t$ later. We 
therfore have to use a comoving frame, which means that the streaming
part of the equation for $f_F$ has the form $\dot{q} \frac{\partial}{\partial
q} f_F(q,t)$.  A  term analogous to $f_B
\frac{\partial\dot{q}}{\partial q}$ cannot
appear in a comoving frame since it counts the difference of
ingoing and outgoing particles in a {\em fixed} phase space volume
element.

The solution of this equation can be obtained in two and three dimensions as
a gradient expansion (see Eq. \ref{timegrad}). 
The solution strategy is completely analogous
to that used for solving the ELBE in two and
three dimensions.  For two dimensions we obtain
\begin{eqnarray}\label{time2d}
\tilde{\psi}^F_0(\tau) &=& \nu e^{-\nu \tau} \nonumber \\
\tilde{\psi}^F_1(\tau) &=& (\frac{1}{4} - \nu \tau) e^{-\nu \tau} \nonumber \\
\tilde{\psi}^F_2(\tau) &=& \frac{1}{\nu} (-\frac{
1}{2} \nu \tau + \frac{1}{4} (\nu
\tau)^2 )e^{-\nu \tau}.
\end{eqnarray}

In 3 dimensions we obtain

\begin{eqnarray} \label{time3d}
\tilde{\psi}^F_0(\tau) &=& \nu e^{-\nu \tau}\nonumber\\
\tilde{\psi}^F_1(\tau) &=& -\nu \tau e^{-\nu \tau} \nonumber \\
\tilde{\psi}^F_2(\tau) &=& \frac{1}{3 \nu} (- \nu \tau + \frac{1}{2} (\nu
\tau)^2 )e^{-\nu \tau}.
\end{eqnarray}

As stated before 
above 
Eq.\ \ref{4.12} the same results may be obtained by
integrating  $\tilde{f}$
over the variables $z$ and $\gamma$.

\section{The distribution of eigendirections $P(\psi)$}
In order to calculate the maximum Lyapunov exponent of a product of
uncorrelated random $2 \times 2$ 
matrices 
$\prod {\bf } U_i(\tau, \phi,\alpha) $, we must 
take into account the fact that the matrices 
${\bf U}_i$  do not in general commute with each other. In
order to use standard theorems to calculate the largest eigenvalue of
a product of random matrices, we have to determine 
the distribution $P(\psi)$ of the angle $\psi$, generated 
by acting with the 
random matrix ${\bf U}$ on the  unit vector 
$\vec{e}(\psi) = (\cos(\psi),\sin(\psi))$ \cite{cpv}. If this
distribution is not isotropic in $\psi$, 
the proper form of $P(\psi)$ must
be determined from the solution of an appropriate Frobenius--Perron equation.
 
In our case, the Frobenius--Perron equation for this distribution is given by
\begin{equation}\label{perron}
P(\psi) = \int_0^{2 \pi} d \psi' P(\psi') \langle 
\delta (\psi - \psi_{1}(\psi',\phi,\alpha)) \rangle 
\end{equation} 
The angle $\psi_{1}(\psi',\phi,\alpha)$ is implicitly defined by 
\begin{equation} \label{psi'}
 {\bf U} \cdot \vec{e}_{\psi'} = |{\bf U} \cdot \vec{e}_{\psi'}| 
\left(\begin{array}{c} \cos\psi_1 \\ 
\sin\psi_1 
\end{array} \right) 
\end{equation}
where $\phi$, $\alpha$ are the scattering angles. 
The distribution function has to obey the normalization condition 
\begin{equation}\label{normP}
\int_0^{2 \pi}  d \psi P(\psi) = 1 
\end{equation}
The average in Eq. (\ref{perron}) $\langle \dots \rangle$ is in our case 
the average over the 
distribution of times of free flight, scattering angles, velocities and space 
multiplied with the survival probability, as defined in Eq. (\ref{repellmax}). 
Since, contrary to the case of motion in an 
external field \cite{lvbd}, the equation of 
motion for ${\bf {U}}$ is the same for an open system as for an
infinite system when there is no external field,  we can 
use here the expression of  $|{\bf U}\cdot \vec{e}_{\psi}|$ derived 
in I.  
To evaluate the ensemble average in the Eq. (\ref{perron}), we have to 
use the gradient expansions for the survival probability and the distribution 
of times of free flight Eqs. (\ref{gradient}, \ref{ggradient}).     
Since the ensemble average involves an average over the velocities, 
the term proportional to $\vec{v} \vec{\nabla} n$ and all other non scalar 
terms vanish. Therefore the deviation of $P(\psi)$ from its isotropic
value, $(2\pi)^{-1}$ 
can at most be of order $\nabla^2 n$. Then, due to the normalization
condition Eq.  
(\ref{normP}) this deviation must also be be proportional to $\cos( m
\psi)$, where m is  
a positive integer. Consequently the only additional term of order 
$\nabla^2 n$ in the expression for the maximum Lyapunov exponent would come 
from multiplying the zeroth order term (in the gradient of $n$) of 
$\langle(\ln |{\bf U} \cdot \vec{e}(\psi)|\rangle$  
(which is independent of 
$\psi$) with the term of order $\nabla^2 n$ in the expression for $P(\psi)$. 
But this averages to zero due to the factor $\cos (m \psi)$. Thus we
may use an isotropic distribution in the angle $\psi$ when calculating
the largest eigenvalue of the product of random matrices, and we then
obtain Eq. (\ref{4.26}).  

We can understand this result also in a more intuitive way. 
This demonstration proceeds in four steps. 
\noindent

1) An isotropic precollisional distribution of $\psi$ will give rise after 
scattering to anisotropies of order $-\bar{q}^2$. To understand this one 
should note that angles $\psi$ and $\psi \pm \pi$ can be identified. This 
merely amounts to an interchange of reference and tangent trajectory. If 
the contributions of such angle pairs are averaged, the term linear in 
$\vec{v}\cdot \nabla n$ cancels. The sole source for the anisotropy in
this 
case is the anisotropy of the survival probability as a function of 
velocity.\\ 
2) A completely anisotropic precollisional distribution, of form 
$\delta(\psi-\psi_0)$ say, will give rise to a postcollisional distribution
that is isotropic in $\psi$, up to corrections of order $-\bar{q}^2$. This 
is a consequence of the isotropic distribution of the azimuthal scattering 
angle $\alpha$ resulting from the random distribution of scatterers. 
Again, the anisotropy is due to anisotropy of the survival probability and 
the linear term in $\vec{v}\cdot \nabla n$ vanishes due to identification 
of $\psi$ and $\psi \pm \pi$.\\ 
3) If we require that the anisotropic part of the distribution of $\psi$ 
cancels on averaging, so does the isotropic part of the distribution
just after a collision, that is, in an ensemble average in which we put 
together all contributions of ensemble members with a collision at an
equal
position and time.\\ 
4) Now since the precollisional anisotropic distribution itself is of order 
$\bar{q}^2$ according to 1) (and nothing will be added to that just
because 
of the result emerging here), its contribution to the postcollisional 
anisotropy has to be of order $\bar{q}^4$.\\ 
The postcollisional anisotropies resulting from the isotropic part of the 
precollisional distribution are fully accounted for by Eq.\
\ref{repellmax}
and so this expression gives correct results through order $\bar{q}^2$.

\end{appendix}

\end{document}